\documentclass[nofootinbib,superscriptaddress,11pt]{revtex4}

\usepackage{epsfig,amssymb,amsmath,latexsym,hyperref}
\newcommand{\be}{\begin{equation}}
\newcommand{\ee}{\end{equation}}
\newcommand{\bea}{\begin{eqnarray}}
\newcommand{\eea}{\end{eqnarray}}
\newcommand{\no}{\noindent}
\newcommand{\nb}{\nonumber}

\newcommand{\de}{\partial}

\newcommand\X{{\cal X}}

\newcommand\m{\mu}
\newcommand\n{\nu}
\newcommand\g{\gamma}

\newcommand\tr{\text{Tr}}

\newcommand\eV{\text{eV}}

\newcommand\ba{\begin{array}}
\newcommand\ea{\end{array}}

\makeatletter
\def\frontmatter@title@above{\null\addvspace{3\baselineskip}}
\def\frontmatter@title@below{\addvspace{1\baselineskip}}
\def  \frontmatter@authorbelow{\show\pippo\addvspace{10\baselineskip}}
\def\frontmatter@preabstractspace{2\baselineskip}
\def\frontmatter@postabstractspace{2\baselineskip}
\makeatother

\begin{document}
\baselineskip=.8\baselineskip
\parskip.2ex

\title{Stars and (Furry) Black Holes in Lorentz Breaking Massive
  Gravity}

\author{D. Comelli}
\affiliation{INFN, Sezione di Ferrara,  I-35131 Ferrara, Italy\phantom{\rule{0em}{2em}}}
\author{F. Nesti}
\affiliation{Dipartimento di Fisica, Universit\`a di Ferrara, I-44100 Ferrara, Italy}
\author{L. Pilo}
\affiliation{Dipartimento di Fisica, Universit\`a di L'Aquila, I-67010 L'Aquila, Italy, and\\
               INFN, Laboratori Nazionali del Gran Sasso, I-67010 Assergi, Italy}

\begin{abstract}
  \no We study the exact spherically symmetric solutions in a class of
  Lorentz-breaking massive gravity theories, using the
  effective-theory approach where the graviton mass is generated by
  the interaction with a suitable set of St\"uckelberg fields.  We
  find explicitly the exact black-hole solutions which generalizes the
  familiar Schwarzschild one, which shows a nonanalytic hair in the
  form of a powerlike term $r^\g$. For realistic self-gravitating
  bodies, we find interesting features, linked to the effective
  violation of the Gauss law: i) the total gravitational mass
  appearing in the standard $1/r$ term gets a multiplicative
  renormalization proportional to the area of the body itself; ii) the
  magnitude of the powerlike hairy correction is also linked to size
  of the body.  The novel features can be ascribed to the presence of the
  Goldstones fluid turned on by matter inside the body; its equation
  of state approaching that of dark energy near the center.  The
  Goldstones fluid also changes the matter equilibrium pressure,
  leading to an upper limit for the graviton mass, $m\lesssim
  10^{-28\div29}\,\text{eV}$, derived from the largest stable
  gravitational bound states in the Universe.
\end{abstract}



\maketitle 

\newpage


\section{Introduction}

\no
The question whether general relativity (GR) is an isolated theory is
interesting from both the theoretical and phenomenological side. It is
known that one can add to the Einstein-Hilbert Lagrangian a tower of
higher order operators, starting with terms quadratic in the
curvature, which produces corrections $\sim (E/M_{P})^n$, where $M_{P}$
is the Planck mass and $E$ is the typical energy of the process we are
considering. Thus the predictions of GR at scales much larger then the
Planck length $1/M_{P}$ are insensitive to them. Modifying GR at
large distances is a totally different business.  Besides its
theoretical interest, the search for large-distance modified theories
of gravity has been motivated by the evidence for cosmic acceleration
and the consequent revival of the long-standing cosmological constant
problem.  The idea is to look for massivelike deformation of GR
featuring a large-distance (infrared) modification of the Newtonian
gravitational potential and a massive graviton.

It is instructive to start from perturbative gravity by considering a
Lorentz-invariant theory of a massive spin-two field~\cite{PF}. The
resulting theory is plagued by a number of diseases that make it
probably unphysical. First, the modification of the Newtonian
potentials is not continuous in the limit of very small graviton mass
$m$ giving a large correction (25\%) to the light deflection from the
sun that is experimentally excluded~\cite{DIS}.  The discontinuity
manifests itself in the weak-field regime and a possible way to
circumvent the problem at the full nonlinear order was proposed by
Vainshtein in~\cite{Vainshtein}: if in the Fierz-Pauli theory (FP) the
linearized approximation breaks down near the Sun the above mentioned
discrepancy cannot be trusted anymore. He proposed using an improved
perturbative expansion, which has a continuous limit for $m \to 0$.
The relative solution is valid only up to a finite distance, and the
question is then whether this solution can be extended up to infinity
and matched with the Yukawa-like solution valid at large
distances~\cite{DAM}. Recent evidences that this can indeed be
achieved by means of a different weak coupling expansion have been
addressed in~\cite{deffayetlast}.\footnote{It is important to anticipate
  here that also the exact solutions obtained below are free of the
  discontinuity problem and can be obtained~\cite{USsph} by a
  non-canonical weak coupling expansion 
where $h_{0i}^2\sim h_{00}\sim
  h_{ij}\sim\epsilon$, see Section~\ref{int}.}  In addition, the FP
theory is also problematic as an effective theory at the quantum
level.  Regarding FP as a gauge theory where the gauge symmetry is
broken by a explicit mass term $m$, one would expect a cutoff
$\Lambda_2 \sim m g^{-1} = (m M_{P})^{1/2}$; however, the real cutoff
is $\Lambda_5 = (m^4 M_{P})^{1/5}$~\cite{NGS} much lower than
$\Lambda_2$. A would-be Goldstone mode is responsible for the extreme
UV sensitivity of the FP theory, which becomes totally unreliable in
the absence of proper UV completion.  These issues cast a shadow on
the  possibility of realizing a Lorentz-invariant theory of massive
gravity~\cite{VAIN}.

It was recently noted that by allowing Lorentz-breaking mass terms for
the graviton the resulting theory can be physically viable~\cite{RUB},
being free from pathologies such as ghosts or low strong coupling
scales, and still lead to modified gravity.  Since mass terms break
the diffeomorphism invariance anyway, this possibility was analyzed
mainly in a model-independent way, by reintroducing the Goldstone
fields of the broken gauge invariance; and by studying their
dynamics~\cite{NGS,DUB}. For a recent review, see~\cite{RUB-TIN}.
Lorentz-breaking massive gravity was also considered in the framework
of bigravity~\cite{USlett}. In particular, we will be interested in a
special phase of Lorentz-breaking massive gravity that has no
propagating scalar perturbations around flat space. In this phase, the
absence of dangerous instabilities survive in the curved spacetime, as
shown in~\cite{uscurved}.

Besides its consistency, in order to be a viable theory massive
gravity has to pass a number a tests. In GR the Schwarzschild solution
is a benchmark and it is thus crucial to study the impact of a
massive deformation  on it. The main goal of this paper is to study both
analytically and numerically the gravitational field produced in by a
spherically symmetric body. The outline is the following.  After the
quick definition of the theory in the Stueckelberg approach in Sec.
\ref{IR}, in Sec.~\ref{SSS} we find the generalized Schwarzschild
solution for Lorentz-breaking massive gravity, and explicitly obtain
the values of the two integration constants entering the exterior
solution as a function of the mass and the size of the body.  This
result is obtained matching the exterior with the interior solutions
for an object of constant density and using the nonstandard
perturbative expansion.  In Sec.~\ref{pert} we discuss the validity
of perturbation theory and present a numerical analysis supporting our
results also when gravity is strong inside the body. In section~\ref{BH} we
discuss some features of the  black-hole solutions.

\section{IR Modified Theories}
\label{IR}

\no Infrared modified gravity theories with manifest diffeomorphism
invariance can be realized by introducing a set of four St\"uckelberg
fields $\Phi^A$ ($A=1,\ldots, 4$) transforming under a diff $\delta
x^\mu=\xi^\mu(x)$ as simple scalars. Then one can build new geometric
objects, manifestly diff invariant, function of the metric field and
the scalar fields, $g^{\mu \nu} \de_\mu \Phi^A \de_\nu \Phi^B$. A
generic function of these quantities, when expanded around a
background, can give rise to graviton mass terms.  One can require
that expanding around a Minkowski background the resulting mass terms
are Lorentz-invariant. However, as discussed in the introduction,
Lorentz invariant massive gravity is rather problematic; therefore, we
relax this requirement and keep just rotational invariance in the
theory. This approach is suited to describe a generic theory in an
effective fashion, and a general action can be written as
\be
\begin{split}
&S = \int d^4x \sqrt g \left( M_P^2 \, R + {\cal L}_{\text{matt}}
\right)+ S_\text{gold} \,, \\
& S_\text{gold} =   \int d^4x \sqrt g \, m^2 M_P^2 \;  {\cal F}({\cal
  X}, V^i, {\cal S}^{ij}) \,,
\end{split}
\label{act}
\ee
where ${\cal F}$ is a rotationally invariant potential, function of
\be
\begin{split}
&\X =  - g^{\mu \nu} \de_\mu \Phi^0 \de_\nu
\Phi^0 \, , \qquad V^i = -  g^{\mu \nu} \de_\mu \Phi^i \de_\nu
\Phi^0\,,\qquad
{\cal S}^{ij} = -  g^{\mu \nu} \de_\mu \Phi^i \de_\nu
\Phi^j \, .
\end{split}
\ee
The constant $m$ sets the graviton mass scale. Note that the breaking
of Lorentz symmetry is directly built in the action; in
Appendix~\ref{bigrav} we discuss a different approach where the
breaking is dynamical.  Furthermore, additional symmetries of the
Goldstone action can be used to single out a particular phase of
massive gravity \cite{DUB}. In particular taking
\be 
{\cal F} \equiv{\cal F}(\X, W^{ij})\,, 
\ee
where $W^{ij} = {\cal S}^{ij} - \X^{-1} \, V^i \, V^j$, the Goldstone
action is invariant under $\Phi^i \to \Phi^i + \xi^i(\phi^0)$ and it
turns out that, in a flat background, only the massive spin 2 tensor
(2 elicities only) propagates~\cite{DUB}.\footnote{It is quite
  interesting that a non perturbative realization of this well-behaved
  ``phase'' is realized also by a general bimetric
  theory~\cite{USlett}, where Lorentz symmetry is broken by the two
  tensor condensates, in the limit where the second metric is
  decoupled.}  The flat background admitted by the action (\ref{act})
can be parametrized as
\be
\bar g_{\mu \nu} = \eta_{\mu \nu} \, , \qquad \bar \Phi^A = (\,a\, t ,\,  b\,  x^i\, ) \, .
\label{lb}
\ee
The background breaks boosts and preserves rotations, as can be seen
by considering the additional effective background metric $\hbox{$\bar
  g_2$}_{\m\n}=\de_\m\bar\Phi^A
\de_\n\bar\Phi^B\eta_{AB}=\text{diag}(- a^2, \, b^2, \, b^2, \,
b^2\,)$ (see Appendix~\ref{bigrav}).  The condition for the existence
of the flat solution (\ref{lb}) is the vanishing of the background
Goldstones energy momentum tensor [see~(\ref{EMT}) in
Appendix~\ref{agold}].  Due to rotational symmetry of the potential
and of the effective $\hbox{$\bar g_2$}_{\mu \nu}$, this condition
amounts to two independent equations that determine $a$, $b$ in terms
of the parameters entering the potential.  The background breaks
Lorentz when $a \neq b$, but as noted above, the Lorentz breaking is
built into the potential, not only in the background
configuration. This is to be contrasted to theories with
\emph{spontaneous} breaking of Lorentz symmetry, as for instance in
bigravity.\footnote{When $a=b$ there are still two equations to be
  solved, while in bigravity (see appendix~\ref{bigrav}) a Lorentz
  invariant background gives only one equation.  In this respect the
  bigravity approach is preferable.}  Here, it is notable that for a
generic potential function $\cal F$ a flat solution is always present,
regardless of a cosmological constant term in the action.

Because of  the nonzero background of the scalar fields, their fluctuations
with respect to the background, $\pi^A=\Phi^A -\bar \Phi^A$; trasform
under diffs as a Goldstone field, $\delta\pi^A=\xi^\mu(x)\de
\bar\Phi^A/\de x^\mu$.  A choice of coordinates setting the Goldstones
fields $\pi^A=0$ fixes the gauge completely and is sometimes called
the \emph{unitary gauge}: all the dynamics is transferred to the
metric.

\section{Spherically Symmetric Solutions}
\label{SSS}
\no
For the spherically symmetric case, one can always find a set of
coordinates where the metric and the Goldstone fields have the
following form
\be
\begin{split}
&ds^2 = - dt^2 \, J(r) + K(r) \, dr^2 + r^2 \, d \Omega^2 \,, \\
& \Phi^0 = a \,  t + h(r)  \, , \qquad \Phi^i =\frac{  x^i}{r} \, \phi(r) \, .
\end{split}
\label{sph}
\ee

We will first derive the exact vacuum spherically symmetric solution
describing the exterior part of the self-gravitating body, then we look
for the interior part. Finally, the exterior and interior parts of the
solution are matched to find  the two
integration constants, $M$ and $S$m  which account for the
gravitational mass of the body and the size of the powerlike term
in the solution, one of the main novel features compared to GR.

\subsection{Exterior, vacuum, exact solution}

The Einstein equations in vacuum read
\be
E^\mu_\nu = \frac{1}{2 M_P^2} \, {T_{\text{g}}}^\mu_\nu \,,
\ee
where $E^\mu_\nu$ is the Einstein tensor and $ {T_{\text{g}}}^\mu_\nu$
is the Goldstone energy momentum tensor (see Appendix~\ref{agold} for
details).  Some general features can be established independently from
the choice of ${\cal F}$. From the ansatz (\ref{sph}) it is clear that
the Einstein tensor is diagonal and therefore also the Goldstone
energy momentum must be diagonal: $ {T_{\text{g}}}^t_r=0$. This gives
\be
\frac{ h'}{K^7 \X^4} \left[\X^4  {\cal F}_x  \, (JK)^3+\X^2
  {\cal F}_1 \phi '^2 \, (JK)^2-2 \X
  {\cal F}_2 \, \phi'^4 \,JK+3{\cal F}_3 \, 
 \phi'^6 \right]=0 \; 
\label{off}
\ee
where $ {\cal F}_1 \, , {\cal F}_2 \, , {\cal F}_3$ and $ {\cal F}_x$
are defined in Appendix~\ref{agold}, and $\X$ [see~(\ref{xdef})] depends
on $h', J \,, K$.  As a result, two branches arise: in the first the
equation is solved by $h'=0$; in the second, with $h' \neq 0$, the
term in square brackets of (\ref{off}) can be solved for $\X$, or
e.g.\ $\phi'$. As noted already in the bigravity context, the first branch
$h'=0$, leads to equations that are very hard to solve analytically;
therefore, we will concentrate on the second branch.

Then, by using the solution to $ {T_{\text{g}}}^t_r=0$, one remarkably
finds that ${T_{\text{g}}}^t_t - {T_{\text{g}}}^r_r=0$ independently
from the potential. Therefore $E^t_t - E^r_r=0$, which implies that,
as in GR,
\be
  K(r) = \frac{k_0}{J(r)} \,,
\ee
where $k_0$ is an integration constant. It not difficult to see that
the remaining Einstein equations consist in 2 equations for 2 unknown
functions $J(r)$, $h(r)$, that can in principle be solved.

\medskip

To proceed further an explicit form for ${\cal F}$ must be provided.
A quite general choice of ${\cal F}$, inspired to a class of bigravity
theories studied in \cite{USsph}, is
\be
\begin{split}
{\cal F} =&  
\ \bigg[\beta_0\,
+  \beta _1 w_{-1} 
+ \beta _2 \left(w_{-1}^2-w_{-2}\right)
+  \beta _3 \left(w_{-1}^3-3 w_{-2} w_{-1}+2 w_{-3}\right)\bigg]\X^{-1}+\\
&{}+\alpha _0
+  \alpha _1 w_1 
+ \alpha _2 \left(w_1^2-w_2\right)
+  \alpha _3 \left(w_1^3-3 w_2 w_1+2 w_3\right)\,.
\label{pot}
\end{split}
\ee
where $w_n=\tr(W^n)$. The truly remarkable feature of the class of
potentials (\ref{pot}) is that the equation ${T_{\text{g}}}^t_r=0$
admits (still for nonvanishing $h'$) a simple solution with
\be
\label{phisol}
\phi =b \, r \, , \qquad 
\bar \alpha _1-4 \bar\alpha _2+6 \bar\alpha _3-\bar\beta _0+2 \bar\beta _1-2 \bar\beta _2  =0 \, ,
\ee
where we defined $\bar \alpha_n = b^{-2 n}\; \alpha_n ,\; \; \bar
\beta_n = a^{-2} \;b^{2n}\;\beta_n $.  The condition~(\ref{phisol})
should be regarded as an equation determining the Goldstone background
$a$ (or $b$) in terms of the parameters of the potential.  Using
~(\ref{phisol}) in the remaining equations, one obtains $h'^2$ (see
Appendix~\ref{appeq}) and finally the exact ``black-hole'' solution
\be
J(r) = 1 - \frac{2 G M}{r} + \Lambda^2 r^2
+ 2 G \, S \, r^\gamma \,.
\label{finJ}
\ee 
Here $M$, $S$ are two integration constants, while $G = 1/16 \pi
M_P^2$ is the Newton constant and $\Lambda^2\equiv\frac{1}{6} m^2
\left(12 \bar \alpha _3-6 \bar\alpha_2+ \bar \alpha _0 -3 \bar\beta
  _1+12 \bar\beta _2-18 \bar\beta _3\right)$ is an effective
``cosmological constant''. The exponent $\g$ is given by
$\g =-2 \left(2 \bar\alpha _2-6 \bar \alpha _3+\bar\beta _1-2 \bar
  \beta _2\right)/(\bar\alpha _1-4 \bar \alpha _2+6 \bar\alpha _3) $.
This kind of solution was first found in~\cite{USsph} in the context
of bigravity theories.

While the $1/r$ qnd $r^2$ terms are also present in the
Schwarzschild-de Sitter solution, the last one represents a power-law
correction to GR whose size is the new integration constant $S$. It
does not contain the graviton mass scale, therefore this term is
present in the exterior solution also in the limit of vanishing
graviton mass.  In other words the influence of the Goldstone modes
survives even in the limit $m\to0$.

Clearly, one can choose $b$ such that the effective cosmological
constant vanishes, $\Lambda=0$. Then, for $\gamma < 2$, the metric
describes an asymptotically flat space, which is just the flat
solution (\ref{lb}). The condition $\Lambda=0$ together with
(\ref{phisol}) are the two conditions for a flat solution determining
$a$, $b$. The masses of fluctuations around flat space
(see~\cite{RUB}) are
\bea
\!\!\!\!\!\!\!\!\!\!\!&&m_0^2= \frac{3}{4} \left (\bar {\alpha } _ 1 -\! 4 \bar {\alpha } _ 2 +\! 
   6 \bar {\alpha } _ 3 -\! \bar {\beta } _ 1 +\! 4 \bar {\beta } _ 2 -\! 
   6 \bar {\beta } _ 3 \right) m^2\,, \quad 
m_1^2=0  \, , \quad 
m_2^2= \frac{1}{2} \left (\bar {\alpha } _ 1 -\! 
   2 \bar {\alpha } _ 2 +\! \bar {\beta } _ 1 -\! 
   2 \bar {\beta } _ 2 \right) m^2 \, ,\nb\\ 
\!\!\!\!\!\!\!\!\!\!\!&&m_3^2= \frac{1}{4}\left(\bar {\alpha } _ 1 -\! 6 \bar {\alpha } _ 3 -\! \bar {\beta } _ 1 +\! 
 8 \bar {\beta } _ 2 -\! 18 \bar {\beta } _ 3 \right)m^2 \,,\quad
m_4^2=\frac{1}{4}\left(\bar {\alpha } _ 1 -\! 4 \bar {\alpha } _ 2 +\! 6 \bar {\alpha } _ 3 -\! 
 3 \bar {\beta } _ 1 +\! 12 \bar {\beta } _ 2 -\! 18 \bar {\beta } _ 3\right)m^2 \,
\label{masses}
\eea
and since the mass of the (spin-two) graviton is $m_2^2$, the
corresponding combination of coupling constants should be positive.

For $\gamma>-1$ the new term $S\,r^\g$ is dominant over the Newtonian
term $M/r$, and accordingly the total gravitational energy of the
solution, evaluated via the Komar mass integral, is infinite (see
Appendix~\ref{energy}). For $\g<-1$ instead the new term is
subleading and the Komar energy is just the mass $M$.  In the
following we will limit ourselves to the case $\g<-1$.

\medskip

If the metric represents the exterior geometry of a spherical
``star''; the integration constants $M$ and $S$ can be computed in
terms of the star parameters by matching the interior and the exterior
solutions. This will be done in the next section.

\medskip

Having the exact solution, it is interesting to discuss its behavior
at large distances, in the case of asymptotically flat solutions,
where it can be compared to the standard weak-field limit.  Curiously,
while $J$ (and $K$) are of the form $1+\epsilon$, with $\epsilon\sim
1/r$, the expression for $h'$, shows that asymptotically $h\sim
\sqrt{\epsilon}$.  So, the solution does not fit in the standard
democratic weak-field expansion, or in other words the standard
weak-field expansion fails to capture the asymptotic behavior. The
correct weak-field expansion is actually nondemocratic, where some
fields are smaller than others.\footnote{This was first pointed out
  in~\cite{USsph}  in the context of bigravity.}  By a gauge
transformation one can eliminate $h'$ and turn on the $h_{0i}$
asymptotic components of the metric, that again are nondemocratic:
$h_{0i}\sim h_{00}^{1/2}\sim h_{ij}^{1/2}$; this is analogous to the
suggestion of Vainshtein for the Fierz-Pauli (Lorentz-invariant)
theory, where in the vicinity of a source this expansion leads to a
non analytic solution which is continuous in the graviton mass
parameter $m\to0$. In that case however, the non analytic solution is
valid only up to a finite radius. The standard weak-field expansion,
leading to the Yukawa falloff is in turn valid only at larger
distances, but is discontinuous in $m$. It was the aim
of~\cite{deffayetlast} to setup a nondemocratic expansion to match the
large and small distance solutions.  Here, it is remarkable that the
above nondemocratic weak-field expansion is valid at all distances, as
shown by the exact solution. In other words, the theory is always
inside the Vainshtein radius.


\subsection{Inside a -Star- and matching}
\label{int} 

Consider now a region where matter is present in the form of a perfect
fluid with energy density $\rho$ and pressure $p$. Though a constant
density fluid is not realistic, it represents a benchmark and in GR a
great deal of
interesting information on the possible stable gravitational
bounded configuration can be derived.  Different from GR, because of
the presence of the Goldstone fields even for a constant density fluid
the resulting Einstein equations are difficult to solve analytically
(the full system of equations in presence of matter is given in
Appendix~\ref{appeq}) and we have to rely on perturbation theory.
Moreover, even the linearization procedure is not straightforward and
one has to set up the nondemocratic weak-field expansion as discussed
in the last section.

The starting point is the expansion in the weak-field $\epsilon$
parameter
\be 
\begin{aligned}
J &= 1+ \epsilon \, J^{(1)} + \cdots \, , \qquad 
&\phi &= b \, r + \epsilon \, \phi^{(1)}+ \cdots \,, \qquad
&\rho &=  \epsilon \, \rho_0 + \cdots  \, ,\\
K &= 1+ \epsilon\, K^{(1)} + \cdots \, , \qquad 
&h &=  \epsilon^{1/2} \, h^{(1)} + \cdots \, ,\qquad
&p &=  \epsilon^2\,p^{(1)} + \cdots  \, , 
\end{aligned}
\label{lin}
\ee
where $\rho_0$ is the constant density of the fluid. As discussed
above, the different parametric size of $h$ is necessary to have a
consistent expansion of the $t$-$r$ component of Einstein equations.
Diff invariance of the matter action alone leads to the conservation
of the matter EMT. Then, by expanding the metric, the pressure is of
order $\epsilon^2$ and can be neglected at leading order.

The linearized solution contains four integration constants, two are
set to zero by imposing regularity in $r=0$; the last two, together
with $S$ and $M$ are determined by matching the interior and the
exterior solution at the star radius~$R$.  For brevity, we only give
here the solution for the case $\Lambda=0$ and $\gamma <0$.  The final
expressions are
\bea \label{J1}
J^{(1)}&=&\frac{G M_0}{R}\left[ \xi^2 -3 +8\;  \mu^2 R^2 \left(
    \frac{\;\xi^2}{\gamma^2 -\gamma -2} - \frac{3\; \xi^4}{5
      (\gamma-3)(\gamma-4)} + \frac{6\; \xi^{1-\gamma}}{(\gamma+3)(2 \gamma-1)(\gamma+1)(\gamma-2)}\right)
\right];  \nb\\[.3cm]
\Delta^{(1)}&=& \frac{ 3 \;G \;M_0}{R}\;(\xi^2-1);    \qquad 
 \phi^{(1)}=3 \;b\; G\; M_0 \left(
\frac{\xi^{-\gamma } }{(\gamma +1)(\gamma+3)}+\frac{\xi^3}{2\;   ( \gamma +3)}-\frac{\xi}{
    2\;(\gamma +1)}
\right)  ,
\label{in}
\eea 
where $M_0=\frac{4}{3}\,\pi\,R^3\,\rho_0$ is the {\it bare} mass of
the star, $\xi\equiv r/R$ and finally $\Delta\equiv JK$ so that
$\Delta^{(1)}=J^{(1)}+K^{(1)}$.\footnote{For $\gamma=-1, \, -3$,
  eq. (\ref{in}) becomes singular and these cases need a special
  treatment given in appendix \ref{special}. }

The parameter $\mu$ is a new important mass combination defined in
terms of the masses around flat space, (\ref{masses}):
\be 
\mu^2\equiv
m_2^2\, \frac{3\, m_4^4-m_0^2\, (m_2^2-3\, m_3^2)}{m_4^4-m_0^2\,(m_2^2-m_3^2)}.
\ee
Note that in general $\mu^2$ can be positive or negative.  The
function $h$ always appears as $h^{\prime2}$ and is obtained in terms
of $J^{(1)}$, $\phi^{(1)}$ and $\Delta^{(1)}$:
\be
\label{hprime}
 {h'}^{(1)\,2} =  a^2\;\left(
 \frac{\g}{12\;m^2\;r^2} 
 ( \Delta^{(1)}-r \; J^{(1)'} - J^{(1)} )-J^{(1)}+\frac{(2-\g)}{r}  \phi^{(1)}\,
\right) \ee
The last unknown, the pressure $p(r)$, is found by using the matter
EMT conservation, and is expressed in terms of $J$ and an integration
constant $p_0$, fixed uniquely when defining the radius $R$ as the
point where $p(R)=0$.  The pressure is of order $\epsilon^2$ but the
ratio $\frac{p}{\rho_0}$ is of order one, therefore we have
\be\label{pr1}
\frac{p(r)}{\rho_0}=  \left[\frac{J(R)}{J(r)} \right]^{1/2}\!\!-1 
\,\simeq\, \epsilon \, \frac{1}{2}
\left[J^{(1)}(R)-J^{(1)}(r)
\right]+ O(\epsilon)^2  \; .
\ee
By matching $J$, $J'$ at $r=R$ with the exterior vacuum solution
(\ref{finJ}) (with $\Lambda=0$), we find the two exterior integration
constants $M$ and $S$ in terms of the parameters of the star:
\be
 M =M_0\left[1-\frac{8\, \mu^2
   R^2}{5 (\g +1)(\g-2)}\right],
\qquad 
 S   = -\frac{24 \;  \mu^2 \,M_0\; R^{1-\gamma} } 
{(\gamma -4)  (\g+1)(2 \gamma -1)(\g -2)}.
\label{match}
\ee
We thus find that the star acts as a source for the new term,
$S\neq0$, and that the bare mass $M_0$ is renormalized.  For $\g<-1$,
both $S$ and the mass shift $\Delta M=M-M_0$ have the same sign of
$\mu^2$. Thus, for $\mu^2>0$ both corrections are positive, $M>M_0$
and $S>0$, while for $\mu^2<0$ both the corrections are negative and
then $M<M_0$ and $S<0$.

The deviation from GR is measured by $S$ which scales as
$R^{4-\gamma}$; as a result, bigger self-gravitating objects produce
larger deviations.  The difference between the gravitational mass seen
by distant observers and the bare mass, $\Delta M=M-M_0$, can be
traced back to the Goldstones' energy density: we have, using the
equation of motion (EOM),
\be
 {T_{\text{g}}}_{tt} =- \, \frac{S (\gamma +1)}{4 \pi} \, r^{\gamma-2} J(r)
 \, ,
\ee
Thus in the exterior region the energy density of the Goldstones is
positive (for $\m^2>0$ and in the range $\gamma <-1$ we are
considering).  Moreover, having a regular solution in all the
spacetime, the Komar energy can be computed as an integral of the
total energy momentum tensor (matter + Goldstones) over a $t=const$
3-ball of radius $\bar r$:
\be
E_{\bar r}= -2 \int_{t=const} \!d^3x\, \sqrt{h} \left(T_\mu^\nu -
  \frac{1}{2} T \, \delta_\mu^\nu \right)\zeta^\mu n_\nu =M_0 +
\Delta M + C_1 \, m^2 \, R^5 \  \, \rho_0 \, 
   \left(\frac{\bar r}{R}\right)^{\gamma +1}\; ;
\ee
with $C_1$ an irrelevant constant. For $\g<-1$, the energy is
finite and equal to $M$ in the limit $\bar r \to \infty$.
\begin{figure}
\centerline{\includegraphics[width=.48\textwidth]{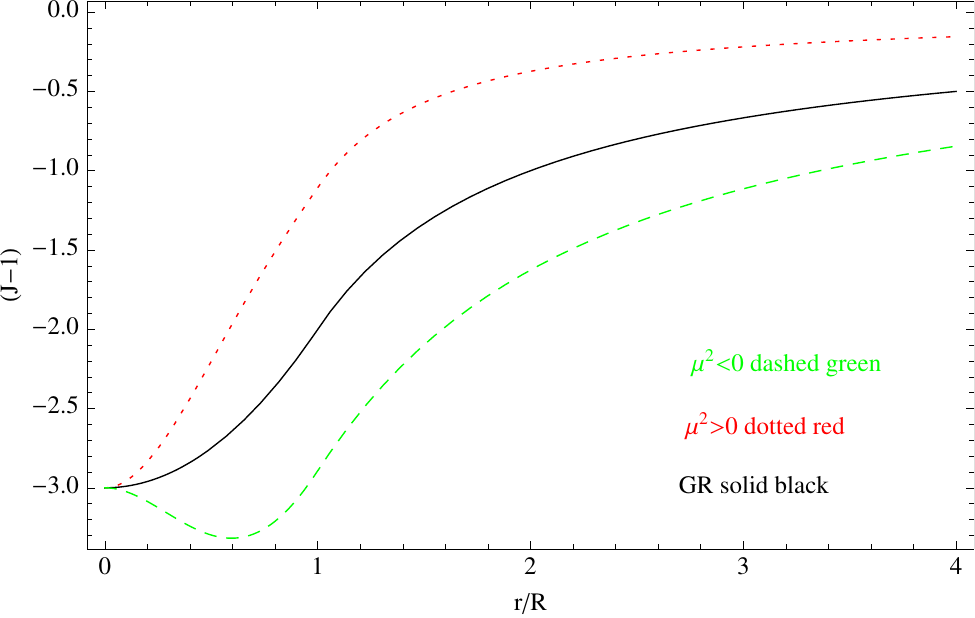}~~~\includegraphics[width=.48\textwidth]{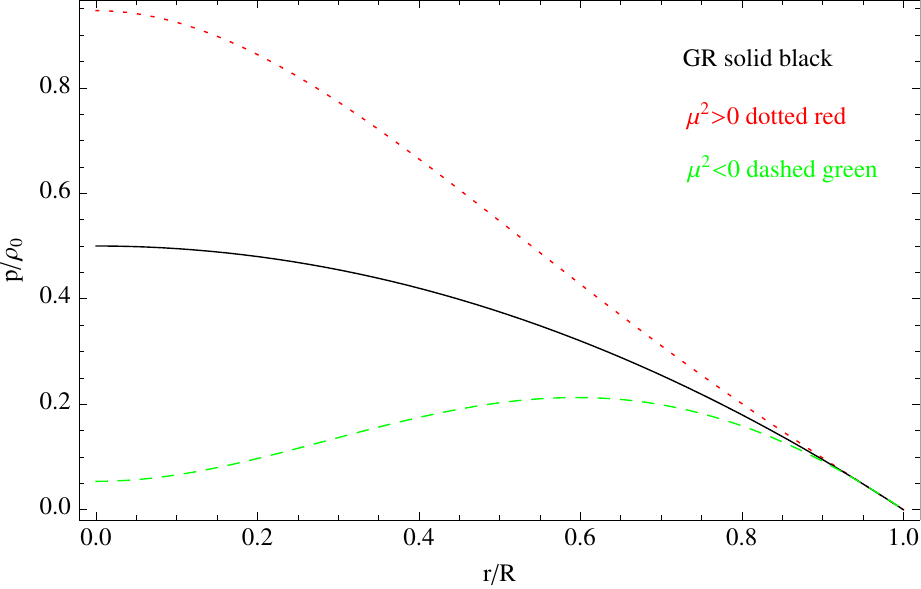}}%
\caption{Plot, in the linear regime, of $J-1$ and $p/\rho_0$ in the
 case of  massive gravity (\ref{J1}) and (\ref{pr1}), both for
  $\mu^2$ positive and negative; also the standard GR case is shown
  for comparison.  The numerical values used are : $\m^2R^2=\pm 3.06$,
  $R_s/R=8.6\times10^{-7}$ and $\g=-2.5$. The units are such that $G
  M_0/R =1$.}
\label{linf}
\end{figure}

Since via $\mu^2$ both $S$ and $\Delta M$ are proportional to the
graviton mass scale $m^2$, we conclude that for $m\to0$ the deviations
from GR disappear, both in the interior linearized solutions and in
the exterior exact one:%
\be
\begin{aligned}
& J^{(1)} \to J^{(1)}_{\text{GR}}\,, \qquad   &&K^{(1)} \to
  K^{(1)}_{\text{GR}}   \qquad &(r \leq R)\\
&J\to J_{\text{GR}}\,,\qquad && K\to K_{\text{GR}}\,,\qquad  M \to M_0 \, , \qquad S \to 0 \,,\qquad &(r\geq R)
\end{aligned}
\label{inm0}
\ee
where $GR$ indicates the corresponding expressions in general
relativity. In the limit of vanishing graviton mass, GR is smoothly
recovered even in the presence of the Goldstone fields $\phi$, $h$
that do not got to zero. Thus, there is no discontinuity for $m \to 0$
in the spherical star solution, and there is no sign of the Vainshtein
scale $r_V$ that in the Pauli-Fierz theory invalidates standard
perturbation theory at scales $r \lesssim r_V$.  Here, the standard
weak-field expansion is never valid, while the nondemocratic
expansion is valid at all distances; in other words, $r_V=\infty$.

The above conclusions apply to realistic weak-field stars like the
Sun. For the strong-field regime the inner star solution can be found
numerically, and we discuss below some of its properties. Of course
the present linearized solution coincides with the exact numerical one
when the density is sufficiently low. In Fig. \ref{linf} the GR
expression of $J$ and $K$ are compared with the ones found here for
massive gravity.

\subsection{Internal pressure and a bound on the graviton mass}
\label{Pbound}

The dimensionless parameter $\mu^2 R^2$ plays a key role in
determining the size of the deviations from GR.  In particular it is
interesting to investigate both the behavior of the internal pressure
and of the total mass $M$, because their renormalization is sensitive
to $\m^2R^2$ and may turn negative when $\mu^2 R^2$ is order one.

\begin{figure}[t]
\centerline{\includegraphics[width=30em]{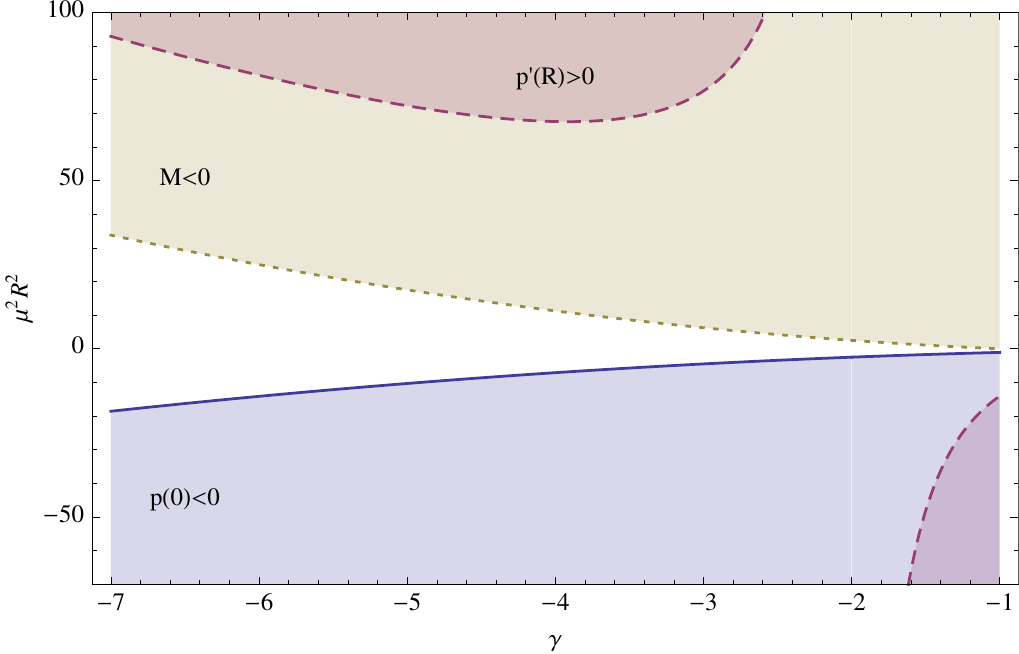}}%
\caption{Excluded regions, in the ($\m^2R^2$--$\g$) plane, from
  positivity of the pressure [$p(0)$ solid line, $p'(R)$ dashed line] and of
  the renormalized mass $M$ (dotted line). The central white region is
  allowed.}
\label{fig:bounds}
\end{figure}

Clearly the stability of the matter system requires the pressure to be
positive. One can see that it is enough to impose this condition at
two regions, near the center and near the surface of the star. In
particular, at the center of the star the requirement of positive
pressure is
\be\label{p0}
\frac{p(0)}{\rho_0} \simeq
\frac{{G M_0}}{2 R}
\left[1-\frac{16\,  \mu^2 R^2 \, (11-2 \g )}{5 \, (2 \gamma -1) 
  (\gamma -4)(\g-2)}\right]  >0 \,,
\ee
while to have positive pressure near the surface, where $p(R)=0$, its
derivative has to be negative:
\be\label{p1R}
\frac{p'(R)}{\rho_0} \simeq
-  
\frac{{G M_0}}{2 R}
\left[1-\frac{16\, \mu^2 R^2 \, ( \g+2 )}
{5 \, (2 \gamma -1)(\g-2) 
  (\gamma -4)}\right] <0 \,.
\ee
On the other hand, the requirement that the renormalized gravitational
mass should be positive is
\be \label{Mren}
M =M_0\left[1-\frac{8\, \mu^2
  R^2}{5 (\g +1)(\g-2)}\right]>0 \, .
\ee
Thus, when $\g<-1$, we have $\delta p(0)\propto \mu^2$, $\delta
p'(R)\propto \mu^2\,\text{sign}(\g+2)$ and $\delta M \propto
-\mu^2$. As a result, for positive $\mu^2$ the bound on $\m^2$ comes
from $p'(R)$ (for $\g<-2$) and $M$, while for negative $\mu^2$ the
bound comes from $p(0)$. This covers the whole range of $\mu^2$.  In
Fig.~\ref{fig:bounds} we show the range of parameters spanned by
such a requirements. The stronger bounds are due essentially to $M$
and $p(0)$.

Assuming no cancellations, or in other words that all mass scales in
the potential are of the same order, the limit on $\m^2$ can be
interpreted as a limit on the graviton mass scale $m^2$.

The typical limit is therefore $ m^2\sim |\mu^2| \leq {\cal O}(1)/ R^2
$; therefore, real stars with e.g.\ $R\sim 10^5\,$km, require $m <
\times 10^{-11}\,$eV.  Considering the Sun, for which $R \simeq 7
\times 10^5\,$Km, and for which the central pressure may not deviate
more than few percents from the standard value~\cite{villante}, we
have $m < \times 10^{-13}\,$eV. However, since the limit does not
depend on the mass of the body but only on its radius, it can be
applied also in the extreme weak-field limit ($R\gg R_s$, see next
section), for instance to very large and rarefied objects. In fact,
larger objects give rise to stronger limits; for instance, the
stability of the largest bound states in the Universe, $R\sim
1\div10\,$Mpc, gives a limit $m<10^{-28\div29}\,$eV.

Though the above considerations are strictly valid for a spherical
symmetric and constant density body, we believe that for more
realistic configuration the above limit can capture the correct order
of magnitude. For other independent limits on graviton mass based on
pulsar timing and CMB polarization effects
see~\cite{massgrav}.\footnote{One should note that the usually quoted
  limits on the graviton mass, like some reported in the PDG, should
  be taken with some care. For instance, the strongest reported
  limit~\cite{lim}, $m< 10^{-32}\,$eV, is derived assuming a
  Yukawa-like potential. Therefore, in the class of theories we are
  considering, such a limit does not apply.}

Let us also comment that the positivity of the renormalized mass $M$
is not strictly required.  In fact, a negative $M$ (and $\g<-1$) would
lead to a repulsive gravity at large $r$ (only) while the body would
still be in equilibrium, having positive pressure.  Such and exotic
behavior of the matter--Goldstones system may deserve further
attention.

\medskip

Regarding the validity of the bounds described above, an important
discussion is in order.  Clearly, the bounds hold, assuming that the
description in terms of gravity coupled to a fluid is valid at the
scales of interest, e.g. galaxies or a cluster of galaxies.  This issue
is far from being trivial, since, e.g. in standard gravity, the
application of the same theory at different scales is guaranteed only
in the Newtonian limit by the linearity of the field equations.  In
nonlinear cases, like for strong field or in the cosmological
evolution, the description in terms of averaged quantities is expected
to need effective corrections, which are under active discussion (see
e.g.~\cite{average}).  In the present theory, two comments can be
made. First, the description of matter in terms of a gravitationally
bound fluid is appropriate, because the typical interparticle distance
is small with respect to the size of the body and to the range of the
gravitational force, which we recall is always infinite[(the Newtonian
term in Eq.~(\ref{finJ})]. On the other hand, the averaging of the
gravitational field equations is non trivial, because, as it is shown
by Eq.~(\ref{hprime}), one of the Goldstone fields enters
quadratically, $(h')^2$. (accordingly, $h'$ is of order
$\sqrt{\epsilon}\sim 1/\sqrt{r}$ in the exact and semilinearized
approach that we described). Because of this nonlinearity, one expects a
violation of superposition of multiple solutions, even in weak-field
regime\footnote{Not to be confused with the violation of Birkoff
  theorem or of Gauss law, which can be violated even in linear
  equations, when departing from purely Newtonian behavior.} Notice
that this nonsuperposition has to arise also in the Lorentz-invariant
version of massive gravity~\cite{Vainshtein}, where one of the fields,
essentially $h'$, enters quadratically and has a $1/\sqrt{r}$
falloff. We expect in fact this to be a generic phenomenon in massive
gravity. What we can conclude is that if one {\em assumes} the present
description of gravity to be applicable at some scale of interest, i.e.
galaxy clusters, then stability of the system leads to the strong
bound derived above.

\section{Perturbation theory and Beyond}
\label{pert}

\subsection{Validity of Perturbation Theory}

\no Let us now study the validity of perturbation theory in both the inner
and outer region of the -star'-.  In GR it is well known that
perturbation theory can be used when $r/R_s \gg 1$, with $R_s = 2 G
M_0$ the Schwarzschild radius.  By inspection of the expression for
the metric perturbation (\ref{in}) inside the body it is clear that
perturbation theory is valid when
\be
R_s \ll R \, , \qquad \m^2 \, R_s \, R \ll 1  \, . 
\label{per1}
\ee
In this section we use $\mu^2$ as $|\mu^2|$ always positive defined.

If we were unable to find an exact solution in the outer region we
could have set up the very same nondemocratic perturbation scheme as
in (\ref{lin}). In fact, solving the linearized equation we find
\be
J^{(1)}= \frac{j_1}{r} + f_1 \, r^2 + j_2 \, r^\gamma + f_2 \,
r^{1-\gamma} \, , \qquad 
K^{(1)}=  - J^{(1)} \, , \qquad  \phi^{(1)}=f_1 \, r + f_2 r^{-
  \gamma} \, ;
\label{out}
\ee 
where $j_{1,2}$ and $f_{1,2}$ are integration constants.  When $\gamma
< -1$, to be in the weak-field regime at large $r$ we have to set
$f_1=f_2=0$, and we get the linearized version of the exterior
solution. The expansion is valid when $j_1/r \ll1$ and $ j_2 \,
r^\gamma \ll1$.  Using the values (\ref{match}) of $j_1$ and $j_2$
obtained from the matching with the interior solution, we get, for
$r>R$, the conditions
\be
r \gg  R_s \, , \qquad r \gg \, R \, \,  (\m^2 \,R_s \,  R)^{1/|\gamma|} \, . 
\ee 
The first condition makes sure that we are away from the would-be
horizon at $r_G$ and it is trivially satisfied in the presence of the
interior part of the solution with $R >R_s$. Notice that in the range
$\gamma < -1$, thanks to (\ref{per1}), also the second condition is
automatically satisfied.  
Then, once perturbation theory is valid in the interior of the body it
can also be used in the exterior part.  When (\ref{per1}) is not
satisfied, the determination of $S$ and $M$ cannot be done using
perturbation theory. In this case one can solve numerically the
equations in the interior part and match the numerical solution with
analytical exact solution that we have found in vacuum.

It is worth stressing  that even within the perturbative region (\ref{per1})
sizable corrections to the gravitational mass $M$ are possible.
This is the case for large bodies with  $R$ of order ${\cal O}(1)/\m$ .
In this case,  the second relation of  (\ref{per1}) becomes equivalent
to the first one $R_s\ll R$,  but the dimensionless quantity $\m^2\,R^2$ 
appearing in $M$ can be sizable.


\subsection{Beyond Perturbation Theory}

\begin{figure}[t]%
\begin{minipage}[b]{0.47\textwidth}
\centering
\includegraphics[width=0.9\textwidth]{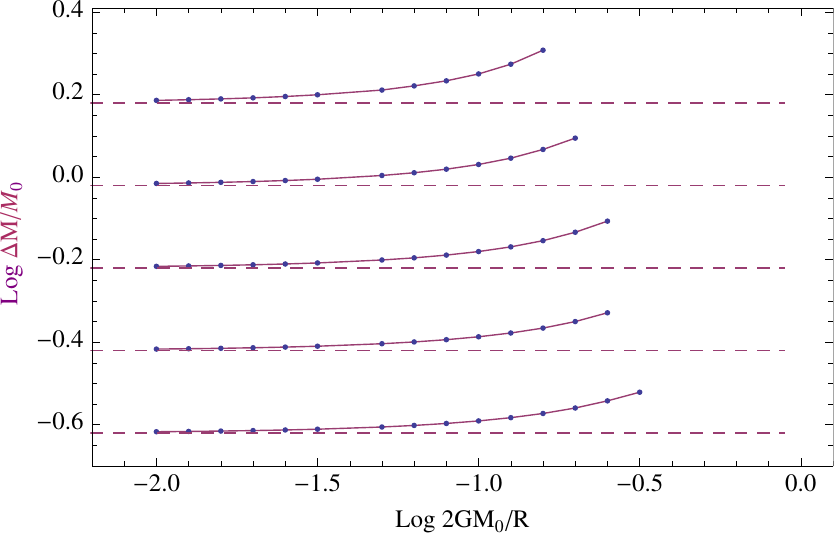}
\end{minipage}
\hspace{0.04\textwidth}
\begin{minipage}[b]{0.47\textwidth}
\centering
\includegraphics[width=0.9\textwidth]{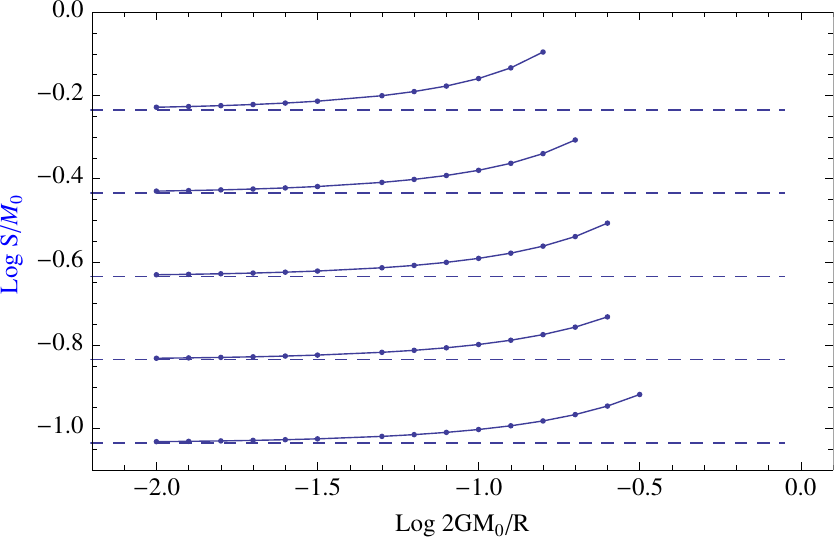}
\end{minipage}%
\caption{Numerical results for $\Delta M=M-M_0$ (left) and $S$ (right)
  for different values of the matter density and of the graviton mass
  scale $\m^2R^2=0.1,\ldots,1$ (lower to upper). The dashed lines
  correspond to the perturbative values; (\ref{match}).}
\label{MS}
\end{figure}

When the equations in (\ref{per1}) are violated, perturbation theory cannot be trusted
anymore and one needs a different tool to see what happens in this
regime. We have solved numerically the Einstein equations in the
interior, still modeling matter as perfect incompressible fluid.
Figure \ref{MS} shows $M$ and $S$ computed numerically for different
values of the graviton mass and matter density.  From the numerical
analysis it is clear that the linearized matching captures the basic
features of the geometry. Indeed, the difference with the linear
predictions are rather small and below 15 \% for $R_s/R\sim 0.5$.

The numerical analysis confirms the absence of a discontinuity for
vanishing graviton mass, as discussed above in the perturbative
analysis [see (\ref{inm0})].  
%
%
Lowering the values of $m$, while keeping fixed the density, we have
found a behavior compatible with $S,\Delta M \to 0 $, up to the
moderate strong-field regime $R_s/R\sim 0.3$, and down to small
graviton mass scale $mR\sim 10^{-4}$.  The numerical analysis thus
shows no sign of discontinuity in $m$.

\begin{figure}[h]
\hskip -.3cm
\centering 
 \includegraphics[width=.5\textwidth]{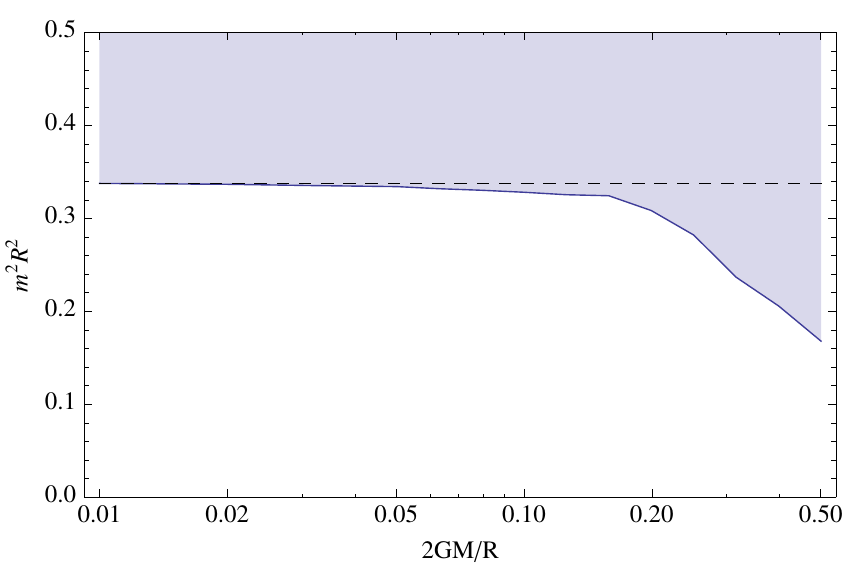}
 \caption{Numerical exclusion plot of the graviton mass [from
   $p(0)>0$] toward the strong field. Here $\g$ is fixed and the
   perturbative limit from (\ref{p0}) is $\m^2R^2 <0.338$ (dashed
   line).}
 \label{bound}
\end{figure}

The remarkable feature that larger bodies gravitate more survives at
the nonperturbative level. The relative mass renormalization for a
body of size $R \sim m^{-1}$ is large, $\Delta M/ M_0 \sim 1$.  In
Fig.~\ref{bound} we extend the perturbative bound on the
dimensionless combination $mR$ to the region where $R_s/R\sim
0.5$. The result is that toward a strong field, the Goldstone pressure
becomes even more negative, pushing the limit on $mR$ even lower.  Of
course real heavy objects are always small, therefore the bound given
above in Sec.~\ref{Pbound} for large weak-field objects is more
stringent.

Finally, the fact that the central pressure receives a negative
contribution from the Goldstone EMT; points to the possibility of
relaxing the Chandrasekar bounds on the density which, we recall,
arise in the strong-field regime when the central pressure
diverges. There may be then a regime where the negative pressure of
the Goldstones balances the high gravitational force, allowing for
compact heavy objects with arbitrarily high mass. We could not reach
this regime with our numerical analysis that becomes arduously
difficult near the origin, when going beyond $R_s/R\sim 0.5$.

\section{Massive Gravity Black Holes with Hair}
\label{BH}

\no It is known that in GR the gravitational hydrostatic equilibrium
cannot be attained for arbitrary $M$ and $R$, for instance a body of
uniform density cannot have a mass larger than $M = 4 R /9$, and for a
fixed radius, this limit applies to any equation of state. The limit is
crucial, because when a star in its late evolutionary state has
exceeded it, the gravitational collapse cannot be stopped and a black
hole is expected to form. 

In GR, the spherically symmetric (uncharged) black holes are
parametrized by its mass $M$. In massive gravity, as we showed, they
are parametrized by an additional parameter, $S$.  Suppose then that
the late stage of the evolution of a star is described by our exterior
metric
\be 
ds^2 = -J \, dt^2 + K \, dr^2 + r^2 d \Omega^2 \, , \qquad  J =1 -
\frac{2 G M}{r} + 2 G S \,  r^\gamma =K^{-1} \,.
\label{bh}
\ee
and for definiteness let us consider here the case with $M>0$. In
general, the hyper-surface at $r=r_H$ where $J(r_H)=0$ is a Killing
horizon: at $r=r_H$ the norm of the timelike Killing vector $\de/\de
t$ vanishes.  If $J$ is never vanishing the singularity at $r=0$ is
naked.  This happens when $S$ is large and positive, $S \geq
\frac{R_s^{|\gamma|}}{4G}$. When $S$ is negative $J$ has a single
zero. Finally, there is a range of $S$ where $ J$ has two zeroes or
one, namely $0< S < \frac{R_s^{|\gamma|}}{4G}$.  When two zeros of $J$
are present, typically the outer Killing horizon is an event horizon
for the black hole. For instance, in the case $\gamma = -2$ the
solution is formally the same of a Reisser-Nordstrom black hole. To
disentangle the causal structure in the general case one has to try to
maximally extend the solution found using a set of Schwarzschild-like
coordinates that fail at the outer Killing horizon.  We leave the
detailed study for a future work.

An other evident feature of the black-hole solution (\ref{bh}) is that
it is one of the few examples of exact black holes with hairs in four
spacetime dimensions.  Their existence is considered to be forbidden
in various classes of theories, by  ``no-hair'' theorems (starting from
\cite{bekenstein}). These are usually based on the positivity of suitable
volume integrals, and often rely on linearity of the field equations.
For instance, a no-hair theorem for the Fierz-Pauli massive gravity
was described already in~\cite{bekensteinPF}, where only the quadratic
FP mass terms were included. As we have proven here, the full theory
of nonlinearly interacting massive gravity on the contrary generates
hairs, extending in the asymptotic region.  These hairs, depending on
the value of $\g$, can even dominate over the Newtonian term, and even
make the solution not asymptotically flat ($\g\geq2$), leading to a type
of \emph{furry} black hole.

The situation is thus similar to known cases of black holes with hairs
in non-Abelian gauge theories (colored black holes). However,
classical non-Abelian fields are actually not observable (due to
confinement). The example provided here would be the first example of
classical hairs from nonlinearly interacting fields.

In the literature there are a number of no-hair theorems which also
consider specifically the presence of extra scalar fields (see for
instance~\cite{Her}) that, in principle apply to our case.  However,
``technical'' simplifying assumptions on the form of scalar field
Lagrangian are often made, rendering the above results not directly
relevant to our nonminimally coupled and interacting set of scalars.

We stress that the fields associated with these hairs (both the metric
and the Goldstones) are nonsingular in all the space, except at the
origin, where in any case the singularity is expected. Also, in the
smooth star solution that we constructed, these fields are never
singular.

Clearly, whether hairs will actually be present in real black holes,
i.e.\ whether $S$ will be nonzero at the final stage of collapse,
will depend on the collapse history, which is surely worth of a
separate study.  For instance, the fact that in the star solution the
corrections to GR are proportional to the star's surface \emph{area},
i.e.\ $\Delta M,S\sim R^2$ (but only in weak-field) may lead one to
think that $S\to 0$ during collapse.  Or on the contrary one may think
that the collapse itself may be stopped or reversed leading to a final
``remnant'' with a non-zero $S$.  We stress in fact that the Birkhoff
theorem is violated in this theory, and therefore the real dynamics is
only constrained by the total energy (not just the Komar integral).  A
related important issue is whether the exact solution is stable under
perturbations: if on one hand the non positivity of the Goldstone
potential $\mathcal F$ hints at instability, on the other hand we
recall that at least in perturbation theory, due to Lorentz breaking,
there is no badly behaving propagating mode.  We leave the study of
stability of the black hole and of the possible scenarios of its time
evolution for a separate analysis.

Finally, the classical hairs, as realized in our solution, seem to
probe the structure of the black hole inside the horizon, and thus may
alter already at classical level its thermodynamical properties.  The
power-law hair $Sr^\gamma$, extending toward the singularity, may also
be sensible to quantum gravity effects, and thus be a probe of the
scales of UV completion.\footnote{Such considerations have also been
  put forward in~\cite{Jac}, in connection with the Lorentz
  breaking. In that case however, the focus is on states propagating
  with different speed of light, thus probing the interior of the
  horizon; here, we remind that there are no other propagating states
  beyond the ordinary graviton. See also~\cite{bumpy} in the case of
  the Lorentz breaking ghost condensate, and~\cite{bebrinst} for
  consideration of LB theories with instantaneous fields.}

\section{Conclusions}  

\no One of the difficulties of massive gravity theories is that
perturbation theory can be very tricky, making hard to  extract
solid phenomenological predictions. This is why it is crucial to find
exact solutions at least in highly symmetric configurations.  The
extension of the familiar Schwarzschild solution is a step forward for
testing massive gravity.

In this work we have addressed this problem in a wide class of
promising massive gravity theories, where the only propagating degree
of freedom is a single massive graviton~\cite{DUB}.  In the
St\"uckelberg spirit, general covariance is restored by a set of
suitable (Goldstone) fields, which are nonlinearly interacting with
the metric field.

We have determined an exact class of black-hole solutions describing
the exterior gravitational field of spherically symmetric compact
bodies, and described its matching with an interior part, describing
the structure of a self-gravitating body. This solution was found
analytically by a suitable nondemocratic weak-field expansion.

For the exterior part of the solution, we have found that
$g_{tt}=g_{rr}^{-1}$ as in GR. Incidentally this implies that the PPN
parameter, which measures the difference between the gravitational
potential felt by a massless and a massive particles, vanishes
identically, and as a result the light-bending in LB massive gravity
coincides with the one of GR.

The differences with respect to GR appears inside $g_{tt}$, where in
addition to the Schwarzschild term, a new nonanalytic one appears,
$g_{tt}=1-2\,G\,M/r+2\,G\,S \,r^\g$, where $\g$ is a constant that
depends on the coupling constants of the theory and $S$ is a new free
integration constant.  The new term $S\,r^\g$ represents a genuine
black-hole hair due to the interacting Goldstone fields.  Depending on
the value of $S$ there are a number of possible scenarios ranging from
a naked singularity at the origin ($S$ large and positive), to a
strong gravity at large distance (for $\g>0$), or to a standard
asymptotically flat behavior with a finite Komar mass equal to $M$,
when $\g<-1$.  We stress that the phase of Lorentz broken massive
gravity we are considering, the range of the static gravitational
potential is infinite even in the presence of massive deformation.

For a realistic spherical body like a star, the Schwarzschild mass $M$
and the new parameter $S$ can be computed (in the physical case
$\g<-1$) by matching the exterior (exact) solution with the interior
(weak-field) one.  This leads to two results:
\begin{itemize}
\item The usual ``bare'' mass of the body $M_0$ is renormalized,
  $M=M_0(1- c_0 \, \m^2R^2)$, where $R$ is the radius of the
  star and  $c_0$ is some numerical factor.
\item $S$ is turned on by the presence of matter, $S\sim M_0\, \m^2 \,R^{1-\g}$. 
\end{itemize}
Here $\m^2$ is a combination of graviton mass scales, which in
principle may be negative.  The sign of $\mu^2$ controls to what
extent the -bare- mass gravitates as seen by a distant observer:
when $\mu^2 <0$, $M$ is larger than the `bare' mass; on the other hand
when $\mu^2 >0$ the body ``degravitates''.  The fact that the
gravitational field depends on the shape (size) of the body is a
signal of the violation of the Gauss law, produced by the nonlinear
interacting Goldstone fields. In both cases it is striking that
$\Delta M=M-M_0$ is proportional to the body surface $R^2$; thus,
larger bodies (de-)gravitate more.  Surprisingly enough,
degravitation can be so large that $M < 0$ while the body internal
structure is still in equilibrium.

Also the internal pressure of the body is subject to a similar
renormalization. By applying this theory of gravitation to known
self-gravitating structures, one can then derive a bound on $\m^2R^2$
by requiring positivity of the pressure, or in other words by
requiring their equilibrium.  Normal stars like the Sun pose very
loose constraints on $\m^2$, but the effect is stronger for larger
objects, so from the largest (and less dense) bound states in the
Universe one may pose a strong limit of order
$\m<10^{-28\div29}\,\eV$. If the various Lorentz-breaking masses are
to be of the same order, this translates into a strong constraint on
the overall graviton mass scale.

The renormalization of mass and pressure, as well as the hair $S
\,r^\g$ are directly produced by the presence of the fluid made of
Goldstone fields.  However, the Goldstone fluid is seeded only by the
presence of matter (at least in weak-field), and the effects disappear
for $M_0\to0$.  In other words, there is no smooth spherical body made
of Goldstone fluid only.  Similarly, the deviations from GR disappear
for vanishing graviton mass scales, $m^2\sim\m^2\to0$, showing no sign
of discontinuity.  This result is exact in the (nondemocratic) weak
field limit, but a numerical investigation of the solutions in
strong-field conditions was 
performed. Up to $2GM/R\sim 0.5$, the analysis
confirmed the absence of discontinuity both in the limit of vanishing
matter or vanishing graviton masses.

\medskip

We did not address the difficult problem of the stability of the
proposed solutions, which is very interesting in view of possible
phenomenological applications, and also in view of the current
conjecture of the instability of configurations with
hairs~\cite{vol}. Notice that because the Birkhoff theorem does not
hold for the gravity modification we have considered, the study of
stability is rather complicated.

Finally, let us also emphasize that  an analysis of the
propagating modes beyond the linearized order is still missing and
should be addressed for these theories to be considered viable. We
leave these important studies for further work.

\bigskip

\no {\bf Acknowledgments.} During this work D.C. was partially
supported by the EU Contract No. FP6 and the Marie Curie Research and  Training Network
-UniverseNet- (MRTN-CT-2006-035863).

\bigskip
\no {\bf Note Added.} The authors of \cite{TI} consider the same
problem of finding a spherically symmetric solution in
Lorentz-breaking massive gravity 
for a particular potential.  According to
that paper, $S$ is always zero and $\phi = r$ for $r \leq R$. However,
this case corresponds to a body made of cosmological constant [see
Eq.~(\ref{eq2}) in Appendix~\ref{appeq}] for which $\Delta =1$
even in the interior, and $\phi = r$ everywhere. Of course, this is
not true for a generic and realistic kind of matter, and $S$ can be
different from zero as shown in this paper.

\begin{appendix}

\section{Various approaches to Massive Gravity}
\label{bigrav}

\no Building masslike terms for the graviton requires the
possibility of scalar combinations of metric. This can done in an
elegant way by introducing a metric $G_{AB}$ in a fictitious manifold
$\overline {\cal M}$, the set of four ``scalars'' can be  interpreted as a
mapping of the physical ${\cal M}$ space into the fictitious one, $\Phi:
{\cal M} \to \overline {\cal M}$, in terms of coordinates $x^\mu \to \Phi^A(x)$. 
The metric in the fictitious space then can pushed back to ${\cal M}$
rendering available the basic tool for constructing diff invariants
generalized mass terms
\be
{g_2}_{\mu \nu} = \frac{\de \Phi^A}{\de x^\mu} \frac{\de \Phi^B}{\de x^\nu} \,
G_{AB} \; .
\ee
The new metric $g_2$ transforms as a standard tensor and can be used
to construct nonderivative
interaction terms by introducing
\be
X^\mu_\nu = {g_2}^{\mu \alpha} g_{\alpha \nu} \, , \qquad \tau_n =
\text{Tr} \left( {X^n} \right)= {(X^n)}^\mu_\mu \, .
\ee
A typical interaction term  will be of the form\footnote{One can also
consider combinations involving $y_n = \text{Tr}(Y^n), \, Y = X^{-1}$.} 
\be
S_{int} = m^2 M_P^2\int d^4x \, \sqrt{g} \, V(\tau_1, \, \tau_2, \, \tau_3 , \, \tau_4) 
\ee
The metric $G_{AB}$ may or may not be considered as dynamical
variable. When $G$ is dynamical, the full action is invariant under
two separate diffs $\text{\it Diff}_1 \times \text{\it Diff}_2$
corresponding to the physical space ${\cal M}$ and to the space
$\overline{\cal M}$. If $G$ is frozen to some background value, the
invariance is broken down to the set of diagonal diff $\text{\it
  Diff}_d$.  The pointwise identification of the two manifold is
obtained imposing $\Phi^A(x) = \delta^A_\mu$, the so called unitary
gauge (UG). In the UG, diff invariance is broken down to $\text{\it
  Diff}_d$ when $G$ is dynamical or completely broken when it is a
frozen background.  As an example, taking $V = \tau_1^2 - \tau_2 -6
\tau_1 +c_0$, in the UG, setting $g_{\mu \nu}= \eta_{\mu, \nu}+ h_{\mu
  \nu}$ and $G_{AB}=\eta_{AB}$, expanding up to the quadratic order
one gets the Pauli-Fierz model.

\section{Goldstone Energy Momentum tensor}
\label{agold}

\no The Goldstone energy momentum tensor for a generic ${\cal F}({\cal
  X},W^{ij})$ is given by
\be
\begin{split}
{T_{\text{g}}}_{\mu \nu} &= m^2M_P^2 \Bigg\{ {\cal F} \, g_{\mu \nu} + 2
{\cal F}_x \,  Y_{\m\n}^{00} +2  
 {\cal F}_{ij}  \left[ \X^{-2} \, V^i \, V^j \,  Y_{\m\n}^{00} 
-   \X^{-1} 
   \left(  V^i Y_{(\m\n)}^{j0}  + V^j 
      Y_{(\m\n)}^{i0}  \right) +   Y_{(\m\n)}^{ij} \right] \Bigg\}
\end{split}
\label{EMT}
\ee
with $Y_{\m\n}^{AB}\equiv  \de_\m \Phi^A \, \de_\n \Phi^B$ and
\be
{\cal F}_x\equiv \frac{\de {\cal F}}{\de \X} \, , \qquad 
{\cal F}_{ij}\equiv \frac{\de {\cal F}}{\de W^{ij}}\,.
\ee
Since $ {\cal F}_{ij}$ a $3\times3$ matrix that depends on $ W^{ij}$,
it can always be written as
\be
 {\cal F}_{ij} =  {\cal F}_1 \, \delta^{ij} +  {\cal F}_2 \, W^{ij} +
 {\cal F}_3 \, {(W^2)}^{ij}  \,,
\ee
with $ {\cal F}_1 \, , {\cal F}_2 \, , {\cal F}_3$ as the scalar
coefficients that depend on the explicit form of $ {\cal F}$.

\section{Einstein equations}
\label{appeq}

\no We give here the full set of Einstein and conservation equations.

The matter EMT conservation gives a first order differential equation
for the pressure\footnote{The same equation is also valid in GR, where
  in addition also $J'/J$ can be eliminated using the $rr$ Einstein equation
  leading to the Oppenheimer-Volkoff equation, which generalizes the equation of 
  hydrostatic equilibrium.}
\be
\label{eq1}
p'+
\frac{J' \left(p+\rho \right)}{2 J}=0 \, .
\ee

The off-diagonal component $tr$, simplifies considerably for the class
of potentials considered and, when $h'\neq 0$ and $m\neq 0$, gives an
equation that can be solved for $\phi^{\prime2}$:
\be \phi^{\prime 2} = \frac{r^4 \mathit{b}^6 \Delta \left(2 r^4
    \mathit{b}^4 \beta _2-2 r^2 \mathit{b}^2 \beta _1 \phi ^2+\beta _0
    \phi ^4\right)}{r^4 \mathit{b}^4 \alpha _1 \phi ^4-4 r^2
  \mathit{b}^2 \alpha _2 \phi ^6+6 \alpha _3 \phi ^8} \ee

Even independently from the potential, using the $tr$ equations it
turns out that the difference between the $tt$ and $rr$ components of
Einstein equations becomes
\be
\label{eq2}
\frac{ \Delta '}{ \Delta^2}=\frac{r\;(p+\rho)}{2\, M_P^2\;J} \, .
\ee
We note that (\ref{eq2}) holds also in GR.  It is interesting to point
out that in the presence of matter in general $\Delta=J K \neq 1$, unless $p
+ \rho =0$ that corresponds to the vacuum or to a body made of
cosmological constant. When $p + \rho =0$, like in the exterior part
of the solution, $\phi = b\, r$ and the Goldstones' EMT is rather simple:
\be
 {T_g}^\mu_\nu= f(r)\, \delta^\mu_\nu \, , \qquad f(r) = \frac{2
   \left(\alpha _1-4 \alpha _2+6 \alpha _3\right) m^2 M_P^2
   \left[a^2 J-a^2+J^2 \left(h'\right)^2\right]}{a^2-J^2
   \left(h'\right)^2}\, .
\ee

The $tt$ component of the Einstein equations depends on the particular
structure of the potential. Its general form can be given as
\be
-\frac{1}{\Delta} \left(\frac{J'}{r}+\frac{J}{r^2}\right)+\frac{p}{2 M_{P}^2}+\frac{1}{r^2}+
\frac{m^2 \Delta }{\left(\phi '\right)^2}\left(H(\phi
  )+\frac{L(\phi )}{{\cal X}}\right) =0 \, ,
\label{eqtt}
\ee
where $H$ and $L$ depend on ${\cal F}$.  Given  the spherically symmetric
ansatz (\ref{sph}), the explicit expression for $\X$ is (notice that
$\X\propto m^2$)
\be
\X=\frac{K-J\;h'^2}{\Delta} \, ,
\label{xdef}
\ee
and one can solve (\ref{eqtt}) for $(h')^2$. Explicitly for our potentials
\be
\label{hp2}
(h')^2=a^2 J^{-2}\left(1-\frac{J m^2 r^2 \left(\alpha _1-4 \alpha _2+6 \alpha
        _3\right)}{r \left(J'(r)+m^2 r \left(\alpha _1-4 \alpha _2+6
          \alpha _3\right)\right)+J-1}\right).
\ee

Finally,  the $\theta\theta$ component of Einstein equations  can be written as
\be
\label{eq5}
J''+J'
   \left(\frac{2}{r}-\frac{r
   \Delta  \left(p+\rho \right)}{4 J
   M_P^2}\right)+12 m^2 \Delta 
   \left[\frac{d_1(\phi) }{\X}+d_2(\phi)\right]-\frac{
   \Delta  \left(3 p+\rho \right)}{2 M_P^2}=0  \,,
\ee
where again the functions $d_1$ and $d_2$ depend the choice of ${\cal F}$.

\section{Behavior of the solutions in the limit $m\to0$}

\no An anomalous behavior of the solutions in the limit $m\to0$ can take
place due to the form of some of the EOM's
\be
m^2 f(x,y)=0\,,\qquad \text{or}\qquad
\begin{cases}
x= m^2 \,g(y)\\
y=\frac{m^2}{x}\,k(x,y)
 \end{cases} 
\ee
where $f,\;g$ and $k$ are smooth functions of their arguments, so that
the solutions obtained for $m=0$ are in general different from the
ones obtained by taking the limit $m\to 0$ of the generic solutions.
Equation ${T_g}_{tr} =0$ is precisely of the first form, while the
$tt$ and $rr$ components of Einstein equations are of the second
form. For the Ssake of compactness in this section all the explicit
formulas refer to the special case: $\alpha_0 = 2, \, \beta_0 =-1, \,
\beta_{i>0} =0, \, \alpha_2 = \alpha_4 =0$.  Assuming $m \neq 0$, we
can solve the equation ${T_g}_{tr} =0$ for ${\phi'}^2$ and Eq.
(\ref{eqtt}) for $\X$; then (\ref{eq5}) becomes
\be
\begin{split}\label{jm}
 J'' & +J' \left[\frac{2 r^3
   (\gamma -2)}{r^4 (\gamma
   -2)-\gamma  \phi
   ^4}-\frac{r \Delta 
   \left(p+\rho _0\right)}{4 J
   M_P^2}\right]+\frac{2 J
   \gamma  \phi ^4}{r^6
   (\gamma -2)-r^2 \gamma 
   \phi ^4}-\\
  &  \frac{12 m^2
   (\gamma -2) \Delta 
   \left(r^6-\phi
   ^6\right)}{r^6 (\gamma
   -2)-r^2 \gamma  \phi
   ^4}+\frac{\Delta 
   \left[\gamma  \phi ^4
   \left(r^2 \left(p+\rho
   _0\right)-4 M_P^2\right)-r^6
   (\gamma -2) \left(3 p+\rho
   _0\right)\right]}{2 M_P^2 r^2
   \left(r^4 (\gamma
   -2)-\gamma  \phi ^4\right)}=0 .
 \end{split}
 \ee
 The GR equations (obtained imposing $m=0$ from the beginning) are
 instead given by eqs (\ref{eq1}) and (\ref{eq2}), and
 \be\label{jgr}
 J''   +J'
   \left(\frac{2}{r}-\frac{r
   \Delta  \left(p+\rho
   _0\right)}{4 J
   M_P^2}\right)-\frac{\Delta 
   \left(3 p+\rho _0\right)}{2
   M_P^2}=0
 \ee
 In vacuum, the difference between Eqs. (\ref{jm}) and (\ref{jgr}) is
 not zero even if we put $m=0$ directly in (\ref{jm}):
\be
\left.(\ref{jm}) - (\ref{jgr}) \right|_{in\,vacuum} =\frac{2 \gamma  \phi ^4}{r^2}\frac{ \left(r J'+J-\Delta \right)
}{r^4
   (\gamma -2)-\gamma  \phi ^4}
\ee
As a result, the equations themselves are discontinuous and we expect
a discontinuity in the space of solutions.

Incidentally, as it is easy to verify, the standard Schwarzschild
solution of GR, i.e. $J=1-2 \,G\,M_0/r$ and $\Delta=1$, satisfies
exactly the above expression. This means that the Schwarzschild
solution is also a solution of the Massive Gravity equations, but
these can have new solutions, corresponding to a nonvanishing
Goldstone EMT.  This is indeed clear in the exact solution given in
the text, where a new term with a new integration constant; $S\,r^\g$;
is present. The discontinuity in $m$ can also be understood from the
exponent $\g$ which is given by a ratio of mass parameters of the
Goldstone potentials, and as such persists in the $m\to 0$ limit.

\medskip

The question is then whether for realistic star solutions the constant
$S$ is vanishing or not for $m\to0$.  In the text we have shown that
in the weak-field regime the solution is smooth in $m$: both $\Delta
M$, $S\to0$ so that the solution reduces to the standard GR one.

In general, inside a medium, the structure of the EOM is the following
\be
E_\mu^\nu = 8 \pi G \, \left( T_\mu^\nu + {T_g}_\mu^\nu \right) \; .
\ee 
The presence of the Goldstones' EMT introduces more equations than in
GR. For instance, in GR both the Einstein tensor and the matter energy
momentum tensor are both diagonal; it is not so when the Goldstones
fields are introduced:
\be
  {T_g}_r^t = -m^2 \, \frac{6  J h' \left(2 r^4 J \, K +\left(r^4 (\gamma -2)-\gamma 
  \phi^4\right) \phi '^2\right)}{r^4 \gamma  \left(K-J
  h'^2\right)^2} \; .
\ee 
The Einstein equation $tr$ leads to $ {T_g}_r^t =0$ that allows us to
solve for $\phi'$ (Of course we assume that neither $J$ nor $h'$ are
vanishing)
\be
{\phi^\prime}^2 = -\frac{2 r^4 J \,  K}{r^4 (\gamma -2)-\gamma
 \phi^4}\,. 
\ee

In GR, introducing  the function $\nu_s(r)$ by
\be
K(r) \equiv \left( 1 - \frac{2 G}{\nu_s(r)} \right)^{-1} 
\ee
the $tt$ equation can be solved in terms of the integral of the matter
energy density. In our case the $tt$ equation is more complicated,
\be
 \nu_s' = 4  \pi  r^2   \rho + \frac{3 r^2  m^2}{G} \left[\frac{(\gamma -2) \phi^2}{r^2 \gamma}-\frac{2 r J }{\gamma  J (r-2 G \mu_s ) {h^\prime}^2-r
   \gamma }-1\right] \; .
\label{tt}
\ee
Notice that again the previous equation has a continuous limit when $m
\to 0$ and reduces to the one of GR unless $J$, $h'$ or $K$ are
singular when $m \to 0$.  The discontinuity turns up when one uses the
$\theta \theta$ equation to eliminate $h'$:
\be
 {h^\prime}^2 = \frac{K}{J} + \, \frac{m^2 \, {\cal H}_n}{m^2
   \, {\cal H}_{d1} + {\cal H}_{d2} } \; ,
\ee
\label{hps}
where ${\cal H}_i$ are suitable functions of $J$, $K$ and $\phi$ that
do not depend explicitly on $m$ (they will depend in general on $m$
implicitly).  Replacing ${h^\prime}^2$ in (\ref{tt}) by using
(\ref{hps}), one finds
\be
\nu_s ' = 4  \pi r^2 \rho  -  \frac{3 \left[2 r^2 \left({\cal H}_{d1} m^2+{\cal H}_{d2}\right) K+{\cal H}_n
  m^2 \left(r^2 \g -(\g -2) \phi^2\right)\right]} {G {\cal
   H}_n  \g } \; .
\label{ttnd}
\ee 
As a result, when one takes the limit $m \to 0$, supposing the $J$,
$K$ and $\phi$ are regular, Eq.\  (\ref{ttnd}) differs from GR if
$\lim_{m \to 0} {\cal H}_{d2} \neq 0$.  At the linearized level, using
(\ref{in}) one finds that $\lim_{m \to 0} {\cal H}_{d2} = 0$,
explaining why the solution is smooth in $m$.

The question whether the strong-field solutions are continuous for
$m\to0$ can at this stage only be addressed numerically. Up to the
field intensity of $R_s/R\sim 0.5$ we could verify that the solution
is actually continuous.

\section{Gravitational Energy}
\label{energy}

\no The gravitation energy can be evaluated by the standard Komar
mass~\cite{KOM}. In the presence of a timelike Killing vector
$\zeta^\mu$ the gravitational energy is given by
\be
E= - \frac{1}{4 \pi G} \int_{\partial \Sigma_t} \sqrt{\rho} d^2 x \, 
n^\mu v^\mu \nabla_\mu \zeta_\nu \,,
\ee
where $\de \Sigma_t$, with unit normal $v^\mu$, is the boundary of the
spacelike 3-surface $t=const$ with unit normal $n^\mu$, finally
$\rho^{\mu \nu}$ is the induced metric in $\de \Sigma_t$. In our case
we take $\de \Sigma_t$ to be the 2-sphere $t=const$: $r=\bar r$ of
large radius $\bar r$ and  the Killing vector is $\zeta = \de/\de
t$. We find
\be
E_{\bar r}= M + S \, \gamma \, \bar r^{\gamma +1}  \, .
\ee
When $\gamma < -1$ we can take the limit $\bar r \to \infty$ and
$E_{\infty }= M$. When $\gamma > -1$ the Komar energy is infinite. A
detailed study of the Hamiltonian approach and in particular of the case with
$\gamma >-1$ will be given elsewhere~\cite{USfut}.

 
 \section{Special cases}
\label{special}

\subsection{The case $\gamma = -1$}
\no In the case with $\g=-1$ we have to change also the exact vacuum solution for the $J$ function:
\be
J=1-\frac{2G\,M}{r}+\frac{2 G \,  S}{r}\log \left(\frac{r}{R} \right)
\ee
the solution of the linearized equations  inside the star of constant density $\rho_0$ read
\be
\begin{split}
& J^{(1)} =\frac{G M_0}{R} \left [\xi^2 -3 +  R^2\bar{\mu}^2_{\g=-1} \,\left(\frac{3}{4}
    \xi ^2 \log \xi  + \xi ^2
 -\frac{1}{8} -\frac{27}{200}  \xi ^4 \right)\right ]\\
&\Delta^{(1)}= \frac{3 G M_0}{R} \left(\xi ^2-1\right)  \\ 
&\phi^{(1)}= \frac{ 3\, b  \, G  \, M_0}{4} \left[\xi^3 - \xi (1+ 2 \log \xi)
  \right],
\end{split}
\ee
where $\xi = r/R$ and $\bar \mu^2_{\g=-1} =( -\alpha _1+16 \alpha
_2-42 \alpha _3+12 \beta _2-36 \beta _3)\;m^2$.  The matching at the
boundary $r=R$ gives the parameters $M$ and $S$ in terms of the star
bare mass and radius
\be\label{s1}
M=M_0\left(1+ \frac{13}{100}\,R^2\bar{\mu}^2_{\g=-1} \right),\qquad
S=-\frac{8}{5}\, M_0 R^2 \bar{\mu}^2_{\g=-1}\, .
\ee 
The (Komar) energy inside a large shell of radius $\bar r$ is given by
\be
E_{\bar r}= M_0\left[ 1  -\frac{12}{25} \, R^2  \mu^2_{\g=-1} 
+ \frac{18}{5} \, R^2\m^2_{\g=-1} \log \left(\frac{\bar r}{R}\right)\right],
\ee
and diverges as a log in the limit $\bar r \to \infty$.

\subsection{The case $\gamma = -3$}
The linearized solution given in the text (\ref{lin}) is modified for
$\g=-3$. The exterior solution has of course the same form, while
the interior solution is drastically different:
\be
\begin{split}
&
J^{(1)} =\frac{G M_0}{R} \left [\xi^2 -3 +  R^2\bar{\mu}_{\g=-3}^2 \left(\frac{531}{1960}
    \xi ^4 - \frac{3}{8} \xi^2  + 
 -\frac{9}{28}  \xi ^4  \log \xi\right)\right ] \\
& \Delta^{(1)}=\frac{3 G M_0}{R} \left(\xi ^2-1\right)  \\
& \phi^{(1)}=\frac{ 3\, b  \, G  \, M_0}{4} \left[\xi + \xi^3 (-1+ 2 \log \xi)
 \right] ,
\label{inl3}
\end{split}
\ee
where $\xi\equiv r/R$ and $\bar \mu^2_{\g=-3} =( -9 \alpha _1+56
\alpha _2-114 \alpha _3+20 \beta _2-60 \beta _3)\,m^2$. The energy is
finite and the parameters of the external solution are given by
\be
M = M_0\left(1 +    \frac{3}{40}\, R^2\bar \mu^2_{\g=-3}  \right)  \, , \qquad
S = \frac{9 }{932}\,M_0  \, R^4\bar \mu^2_{\g=-3} \, .
\ee

\end{appendix}


\begin{thebibliography}{99}


\bibitem{PF} M.~Fierz and W.~Pauli,
\emph{Proc.\ Roy.\ Soc.\ Lond.\ } A {\bf 173}, 211 (1939).

\bibitem{DIS}
H.~van Dam and M.~J.~G.~Veltman,
\emph{Nucl.\ Phys.\ }  B {\bf 22} (1970) 397;\\ 
Y.~Iwasaki,
\emph{Phys.\ Rev.\ }  D {\bf 2} (1970) 2255;\\
V.I.Zakharov, \emph{JETP Lett.} {\bf 12} (1971) 198.

\bibitem{deffayetlast}
 E.~Babichev, C.~Deffayet and R.~Ziour,
  arXiv:1007.4506 [gr-qc].

\bibitem{deRham:2010ik}
  C.~de Rham and G.~Gabadadze,
  Phys.\ Rev.\  D {\bf 82} (2010) 044020 and
  Phys.\ Lett.\  B {\bf 693} (2010) 334.

\bibitem{NGS}
N.~Arkani-Hamed, H.~Georgi and M.~D.~Schwartz,
\emph{Annals Phys.\ } {\bf 305} (2003) 96.
 
\bibitem{Vainshtein}
  A.~I.~Vainshtein,
Phys.\ Lett.\  B {\bf 39}, 393 (1972).

\bibitem{BD}
D.~G.~Boulware and S.~Deser,
\emph{Phys.\ Rev.\ } D {\bf 6} (1972) 3368.

\bibitem{DAM}
T.~Damour, I.~I.~Kogan and A.~Papazoglou,
\emph{Phys.\ Rev.\ } D {\bf 67} (2003) 064009.

\bibitem{VAIN}
  G.~Dvali,
  \emph{New J.\ Phys.}  {\bf 8}, 326 (2006); \\
A.~Vainshtein,
  Surveys High Energ.\ Phys.\  {\bf 20}, 5 (2006);\\
  P. Creminelli, A.  Nicolis, M. Papucci, E. Trincherini, \emph{JHEP} {\bf 09} 
  003.

\bibitem{RUB}
  V.~A.~Rubakov,
  arXiv:hep-th/0407104.

\bibitem{RUB-TIN}
  V.A.~Rubakov, P.G. Tinyakov
 Phys.\ Usp.\  {\bf 51} (2008) 759.

\bibitem{USlett}
  Z.~Berezhiani, D.~Comelli, F.~Nesti and L.~Pilo,
  \emph{Phys.\ Rev.\ Lett.}\  {\bf 99} (2007) 131101.

\bibitem{uscurved}
D.~Blas, D.~Comelli, F.~Nesti and L.~Pilo,
  Phys.\ Rev.\  D {\bf 80} (2009) 044025.







\bibitem{DUB}
S.~L.~Dubovsky,
  \emph{JHEP} {\bf 0410}, 076 (2004).

\bibitem{USsph}
  Z.~Berezhiani, D.~Comelli, F.~Nesti and L.~Pilo,
 JHEP {\bf 0807}, 130 (2008).


\bibitem{TI}
M.~V.~Bebronne and P.~G.~Tinyakov,
  JHEP {\bf 0904} (2009) 100.

   M.~V.~Bebronne,
  arXiv:0910.4066 [gr-qc]  and
  Phys.\ Rev.\  D {\bf 82} (2010) 024020.

\bibitem{KOM}
A.~Komar
  Phys Rev. {\bf 113}, 934 (1959).

\bibitem{USfut}
 D.~Comelli, F.~Nesti and L.~Pilo,
to appear.

\bibitem{massgrav}
S.~L.~Dubovsky, P.~G.~Tinyakov and I.~I.~Tkachev,
  Phys.\ Rev.\ Lett.\  {\bf 94} (2005) 181102;


M.~Pshirkov, A.~Tuntsov and K.~A.~Postnov,
  Phys.\ Rev.\ Lett.\  {\bf 101} (2008) 261101;

S.~Dubovsky, R.~Flauger, A.~Starobinsky and I.~Tkachev,
  Phys.\ Rev.\  D {\bf 81} (2010) 023523.

\bibitem{lim}
  S.~R.~Choudhury, G.~C.~Joshi, S.~Mahajan and B.~H.~J.~McKellar,
 Astropart.\ Phys.\  {\bf 21}, 559 (2004).


\bibitem{villante}
  F.~L.~Villante and B.~Ricci,
  Astrophys.\ J.\  {\bf 714} (2010) 944.

\bibitem{average}
  G.F.R.~Ellis, T.~Buchert,
  Phys.\ Lett.\  {\bf A347 } (2005)  38-46;\\
  R.J.~v.~d.~Hoogen,
  [arXiv:1003.4020 [gr-qc]].

\bibitem{bekenstein}
J. Bekenstein, \emph{Phys.\ Rev.\ } {\bf 5} (1972) 1239.

\bibitem{bekensteinPF}
J. Bekenstein, \emph{Phys.\ Rev.\ } {\bf 5} (1972) 2403.


\bibitem{bumpy}
  S.~Dubovsky, P.~Tinyakov and M.~Zaldarriaga,
  JHEP {\bf 0711} (2007) 083;

\bibitem{bebrinst}
 M.~V.~Bebronne,
  Phys.\ Lett.\  B {\bf 668} (2008) 432
  [arXiv:0806.1167 [gr-qc]].

\bibitem{Jac}
  T.~Jacobson and A.~C.~Wall,
  Found.\ Phys.\  {\bf 40} (2010) 1076.

\bibitem{Her}
  T.~Hertog,
  Phys.\ Rev.\  D {\bf 74} (2006) 084008.
\bibitem{vol}
  D.~V.~Gal'tsov, E.~A.~Davydov and M.~S.~Volkov,
  Phys.\ Lett.\  B {\bf 648}, 249 (2007)
  [arXiv:hep-th/0610183].

\end{thebibliography}
\end{document}